\documentclass[pra, showkeys, reprint, nofootinbib]{revtex4-1}
\usepackage[english]{babel}
\usepackage[utf8]{inputenc}
\usepackage[colorinlistoftodos, color=green!40, prependcaption]{todonotes}
\usepackage{color}
\usepackage{amsthm}
\usepackage{mathtools}
\usepackage{amsmath}
\usepackage{physics} 
\usepackage{xcolor}
\usepackage{graphicx}
\usepackage{soul}
\usepackage{bm}
\usepackage{mathbbol}
\usepackage[pdftex, pdftitle={Article}, pdfauthor={Author}]{hyperref} 

\newtheorem{definition}{Definition}

\begin{document}
\title{Parameter Concentration in Quantum Approximate Optimization} 

\author{V.~Akshay}
    \email[e-mail:]{akshay.vishwanathan@skoltech.ru}
      \homepage{http://quantum.skoltech.ru}
    \affiliation{Skolkovo Institute of Science and Technology, 3 Nobel Street, Moscow, Russia 121205}
\author{D.~Rabinovich}
    \affiliation{Skolkovo Institute of Science and Technology, 3 Nobel Street, Moscow, Russia 121205}
\author{E.~Campos}
    \affiliation{Skolkovo Institute of Science and Technology, 3 Nobel Street, Moscow, Russia 121205}
\author{J.~Biamonte}
    \affiliation{Skolkovo Institute of Science and Technology, 3 Nobel Street, Moscow, Russia 121205}

\date{March 2021}

\begin{abstract}
The quantum approximate optimization algorithm (QAOA) has become a cornerstone of contemporary quantum applications development. In QAOA, a quantum circuit is trained---by repeatedly adjusting circuit parameters---to solve a problem.  Several recent findings have reported parameter concentration effects in QAOA and their presence has become one of folklore: while empirically observed, the concentrations have not been defined and analytical approaches remain scarce, focusing on limiting system and not considering parameter scaling as system size increases. We found that optimal QAOA circuit parameters concentrate as an inverse polynomial in the problem size, providing an optimistic result for improving circuit training. Our results are analytically demonstrated for variational state preparations at $p=1,2$ (corresponding to 2 and 4 tunable parameters respectively). The technique is also applicable for higher depths and the concentration effect is cross verified numerically. 
Parameter concentrations allow for training on a fraction $w < n$ of qubits to assert that these parameters are nearly optimal on $n$ qubits. Clearly this effect has significant practical importance. 



\end{abstract}

\maketitle

\section{Introduction}

Variational quantum algorithms are the centerpiece of study in the theory and application of modern quantum computing algorithms. Such algorithms are designed to alleviate certain systematic limitations of near term devices, such as variability in pulse timing and limited coherence times \cite{harrigan2021quantum,pagano2019quantum, guerreschi2019qaoa,butko2020understanding}, by the use of a quantum to classical feedback loop. In particular, the Quantum Approximate Optimization Algorithm (QAOA) \cite{Farhi2014} was developed to find approximate solutions to combinatiorial optimization problems \cite{niu2019optimizing,Farhi2014,lloyd2018quantum,morales2020universality,Zhou2020,wang2020x,Brady2021,Farhi2016,Akshay2020,Farhi2019a,Wauters2020,Claes2021,Zhou}.
Recent milestones include experimental demonstration of $3p$-QAOA (depth-three, corresponding to six tunable parameters) using twenty three qubits \cite{harrigan2021quantum}, universality results \cite{lloyd2018quantum,morales2020universality}, as well as several results that aid and improve on the original implementation of the algorithm \cite{Zhou2020,wang2020x,Brady2021}. Although QAOA exhibits provable advantages such as recovering a near optimal query complexity in Grover's search \cite{Jiang2017a} and offers a pathway towards quantum advantage \cite{Farhi2016}, limitations are known for low depth QAOA \cite{Akshay2020,hastings2019classical,Bravyi2019}. Higher depth versions may be needed to overcome such limitations. However, exact analysis is scarce and only describes QAOA on specific instances including e.g.~fully connected graphs \cite{Farhi2019a, Wauters2020,Claes2021}. A general analytical approach has remained unknown.                   

Similar to most variational algorithms, QAOA consists of an outer loop classical optimization which assigns parameters to a quantum circuit in order to minimize an objective function. However, this step becomes challenging beyond low depth due to the simultaneous optimization of several parameters. Although layer-wise training, a learning strategy designed to reduce optimization time, has been shown to work via a re-parameterization of search parameters \cite{Zhou}, such strategies become sub-optimal in certain scenarios (abrupt training transitions \cite{campos2020abrupt}). A different approach towards reducing the complexity associated with the classical optimization step in QAOA is by leveraging \textit{concentrations}.

Concentrations arise in the literature as folklore: though mentioned in numerical and even analytical studies \cite{brandao2018fixed,Streif2020,sack2021quantum}, their precise definition, scaling behavior, and analytic prediction is lacking. State of the art analytical approaches were based on the fully connected Sherrington-Kirkpatrick model. For general depth $p$, it was shown that QAOA becomes instance independent in the infinite system size limit ($n \to \infty$) \cite{Farhi2019a}. Although this result applies to concentrations with respect to instances, the scaling behavior of optimal parameters were not addressed. In addition to instance concentrations, several numerical studies report \textit{distributions} over optimal parameters even when QAOA on random instances are considered \cite{Streif2020,Zhou,crooks2018performance}. Furthermore, such distributions are empirically shown to behave non-trivially with respect to~$n$ and therefore add to the folklore of concentrations.

In this work we explicitly define \textit{parameter concentrations} (Section \ref{definition}), an effect where optimal QAOA parameters for a fixed depth ansatz circuit retain optimality independent of the problem size (increasing number of qubits $n$). 
We introduce an analytical approach to describe the behavior of optimal parameters for depth $p=1,2$ QAOA on variational state preparation (Section \ref{description}). From this we recover optimal parameter scaling with respect to the number of qubits and establish parameter concentration. The same analysis is applicable for circuits of arbitrary depth $p\ge3$ and a numerical verification of parameter concentrations was carried out. For depth $p=5$ and up to $n=17$, we recover scaling for optimal parameters which also demonstrate parameter concentration (Section \ref{results}). 

Our definition of parameter concentrations severely restrict the behavior of optimal parameters. If parameters concentrate, one can train $p$ levels on a fraction $w<n$ of qubits and assert these parameters are nearly optimal on $n$ qubits and $p$ levels. This has evident practical significance and is elaborated more in the Discussion \ref{discussion}.

\section{Parameter concentration definition}
\label{definition}

We represent $\ket{\psi(\bm \gamma, \bm \beta)}$ as the variational state generated by a $p$-depth QAOA circuit for real hyperparameters $\bm \beta\in[0,\pi)^{\times p}$, and $\bm \gamma\in[0,2\pi)^{\times p}$. ~Let~$\bm\gamma_n,~\bm \beta _n= \arg\min \bra{\psi(\bm \gamma, \bm \beta)}\mathcal{H}_z\ket{\psi(\bm \gamma, \bm \beta)}$, where $\mathcal{H}_z$ describes a problem instance on $n$ qubits. Note that such a set of parameters $\bm \beta_n =(\beta_1,\dots, \beta_p)$ and $\bm \gamma_n = (\gamma_1 \dots \gamma_p)$ is not necessarily unique. 

\begin{definition}[Parameter Concentration] Parameters concentrate whenever
\begin{align*}
&\exists ~l>0:\forall ~ \bm \beta_{n}, \bm \gamma_{n} ~\exists~ \bm \beta_{n+1}, \bm \gamma_{n+1} :\\
&\abs{\bm\beta_{n+1}-\bm\beta_{n}}^2+\abs{\bm\gamma_{n+1}-\bm\gamma_{n}}^2=\mathcal{O}\left(\dfrac{1}{n^l}\right).
\end{align*}
\end{definition}
In other words, parameter concentration implies that whichever set of optimal parameters one determines for $n$ qubits, there exists at least one set of parameters, polynomialy close (in $n$), which is optimal for $n+1$ qubits.

\section{Variational state preparation}
\label{description}

State preparation has implications on optimal control theory \cite{Li2017,Brif2010,larrouy}, quantum chemistry \cite{Leibfried2012,sugisaki}, many body physics \cite{Verstraete2009,Chiu2018}, and other areas \cite{Saito2006,Gelbwaser-Klimovsky2014,Kardashin2020}. Although adiabatic approaches can prepare an arbitrary state \cite{richerme2013experimental,bernien2017probing}, the same can be addressed with a variational approach, e.g.~variational quantum eigensolver (VQE) or QAOA, by tuning short quantum circuits \cite{Wauters2020,streif2019comparison,ho2018efficient,ho2019,Bravyi2019,Bartschi2020,Kuzmin2020, kandala2017hardware}.

Variational state preparation can be stated as follows: let $\ket{t}$ be a $n$-qubit target state in the computational basis. The task is to variationaly prepare a candidate state with high overlap with  $\ket{t}$. In QAOA the candidate state $\ket{\psi (\bm\gamma, \bm\beta)}$ is parametrized as:
\begin{equation}
    \ket{\psi(\bm\gamma,\bm\beta)} =  \prod\limits_{k=1}^p e^{-i \beta_k \mathcal{H}_{x}} e^{-i \gamma_k \ketbra{t}{t}}\ket{+}^{\otimes{n}},
    \label{ansatz}
\end{equation} 
where $\mathcal{H}_x = \sum_{i=1}^{n} X_{i}$ is  the standard one-body mixer Hamiltonian with $X_i$ being the Pauli matrix, applied to the $i$-th qubit, and $\gamma_k\in[0,2\pi)$, $\beta_k\in[0,\pi)$.

The optimization task is to maximize the overlap between the candidate state $\ket{\psi(\bm\gamma,\bm\beta)}$ and the target state $\ket{t}$ given by
$\abs{\braket{t}{\psi(\bm \gamma, \bm \beta)}}^2$. Note that the problem is equivalent to the minimization of the problem Hamiltonian $\mathcal{H}_{z}  = \mathbb{1} - \ketbra{t}{t}$,
\begin{equation}\label{objfun}
    \min_{\bm\gamma,\bm\beta} \bra{\psi ( \bm\gamma, \bm\beta)} \mathcal{H}_{z} \ket{\psi (\bm\gamma, \bm\beta)} = 1 - \max_{\bm\gamma,\bm\beta} \abs{\braket{t}{\psi(\bm \gamma, \bm \beta)}}^2.
\end{equation}

\section{Results}
\label{results}

In order to calculate optimal parameters, i.e.~paramerets that achieve maximum overlap, we evaluate the conditions for zero gradients. For $p=1$, the two equations for $\gamma\equiv\gamma_1$ and $\beta\equiv\beta_1$ can be further simplified into a single expression in terms of either $\gamma$ or $\beta$. Since no general solution has ever been obtained, approximate solutions are found at $n\gg1$. The technique can be extended for $p \geq 2$. However, for higher depth, the sets of equation for zero gradients may not simplify as in $p=1$. Still, solutions can be found in the limit $n \to \infty$ along with their next order corrections. In both cases, parameters are found to concentrate. 

\subsection{Parameter concentration for $p=1$}

For single depth, the ansatz state (\ref{ansatz}) becomes:
\begin{equation}
    \ket{\psi(\gamma,\beta)} =  e^{-i \beta \mathcal{H}_{x}} e^{-i \gamma \ketbra{t}{t}}\ket{+}^{\otimes{n}}.
\end{equation}
We calculate the amplitude as:  
\begin{align}
    g_1(\gamma,\beta)&= \braket{t}{\psi(\gamma,\beta)}\notag\\
    &=\dfrac{1}{\sqrt{2^n}}
    \big(e^{-i\beta n}+\cos^n \beta(e^{-i\gamma}-1)\big),
    \label{1_layer_amp}
\end{align}
then the overlap becomes:
\begin{align}
     F_1(\gamma,\beta) =\dfrac{1}{2^n} \big[1&+ 2 \cos^{n}{(\beta)}\left(\cos{(\gamma - n\beta)} - \cos{(n\beta)}\right) \nonumber \\
     & + 2\cos^{2n}{(\beta)}\left( 1 - \cos{\gamma} \right) \big].
    \label{overlap}
 \end{align}

We are concerned with parameters that maximize \eqref{overlap} and therefore must satisfy $\partial_\gamma F_1(\gamma,\beta)=\partial_\beta F_1(\gamma,\beta)=0$. From these conditions the following is established:

\begin{align}
    &\tan \gamma = \dfrac{\sin n\beta}{\cos n\beta -\cos^{n}\beta},
    \label{d_gamma}\\
    &\tan \dfrac{\gamma}{2}= \dfrac{\cos(n\beta+\beta)}{2\cos^{n}\beta\sin\beta-\sin(n\beta+\beta)}.
    \label{d_beta}
\end{align}
Merging \eqref{d_gamma} and \eqref{d_beta}
we arrive at:
\begin{equation}
   \cos^n\beta\sin2\beta-\sin(n+2)\beta=0.
  \label{beta_eq}
\end{equation}
Substituting \eqref{beta_eq} into \eqref{d_beta}, we finally relate optimal parameters as $\gamma=\pi-2\beta$.

We now consider solutions for \eqref{beta_eq}. For a given $n$, it can be rewritten as a polynomial equation of power $n$ which does not have solutions in radicals. However, one can see that for $n\to\infty$, $\beta = \dfrac{\pi}{n}+\mathcal{O}(n^{-2})$. We find correction to the asymptotic solution as $\beta=\dfrac{\pi}{n}+\delta\beta$, and calculate $\delta \beta=-\dfrac{4\pi}{n^2}$ up to second order in $1/n$. 
Thus for the optimal parameters:
\begin{eqnarray}
    \beta=\dfrac{\pi}{n}-\dfrac{4\pi}{n^2}+\mathcal{O}(n^{-3}),
    \label{beta}
    \\
    \gamma=\pi-\dfrac{2\pi}{n}+\dfrac{8\pi}{n^2}+\mathcal{O}(n^{-3}).
    \label{gamma}
\end{eqnarray}

\begin{figure}[!tbh]
   \centerline{\includegraphics[clip=true,width=3.2in]{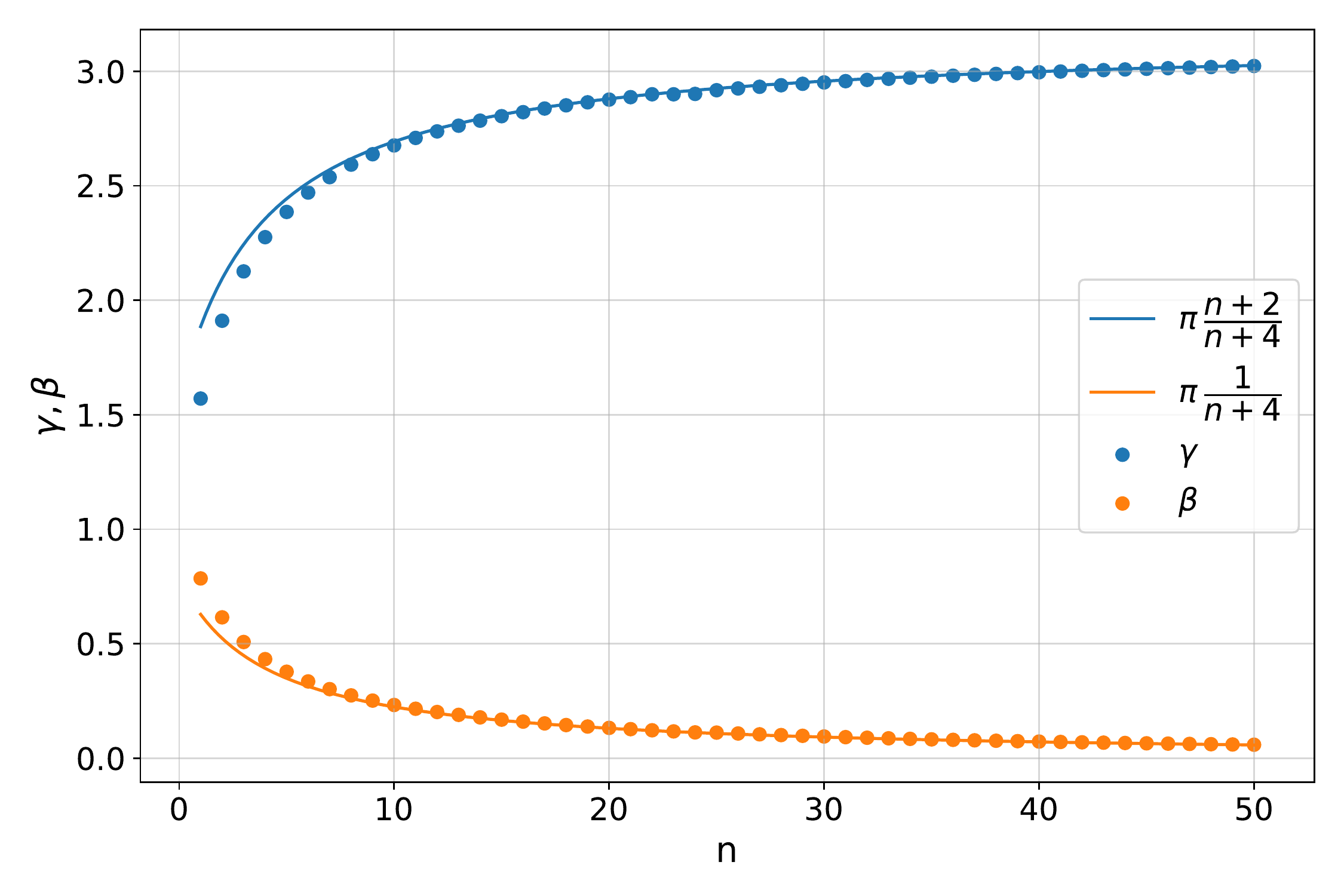}}
       \caption{Comparison of analytical and numerical results for optimal parameters $\beta$ (orange) and $\gamma$ (blue) with respect to the number of qubits $n$. Dots represent numerical solution to the maximization problem \eqref{objfun}.}
 \label{angles}
 \end{figure}

Interestingly, solutions $\beta=\dfrac{\pi}{n+4}$ and $\gamma=\pi-2\beta=\pi\dfrac{n+2}{n+4}$, which turn into  \eqref{beta} and \eqref{gamma} for $n\gg1$, approximate optimal parameters even for small $n$ (see Figure~\ref{angles}).


Additionally, there is a symmetric solution $\beta\to\pi-\beta$ and $\gamma\to2\pi-\gamma$, which delivers the same global maximum to the overlap \eqref{overlap}. For large $n$, \eqref{beta_eq} also admits other solutions. These however, deliver only local maxima and thus are out of interest. Therefore,
\begin{align}
     \abs{\bm\beta_{n+1}-\bm\beta_{n}}^2+\abs{\bm\gamma_{n+1}-\bm\gamma_{n}}^2\approx\nonumber\\ \approx\dfrac{5\pi^2}{(n+4)^2(n+5)^2}=\mathcal{O}\left(\dfrac{1}{n^4}\right),
\end{align}
which establishes parameter concentration for $p=1$.

\subsection{Parameter concentration for $p=2$}

For $p=2$ the ansatz becomes: 
\begin{align}
    \ket{\psi(\gamma_1,\beta_1,\gamma_2,\beta_2)}& =\notag\\
    e^{-i \beta_2 \mathcal{H}_{x}}& e^{-i \gamma_2 \ketbra{t}{t}} e^{-i \beta_1 \mathcal{H}_{x}} e^{-i \gamma_1 \ketbra{t}{t}}\ket{+}^{\otimes{n}}.
\end{align}
The corresponding amplitude can be expressed in terms of the amplitude at $p=1$ from \eqref{1_layer_amp}:
\begin{align}
&g_2(\gamma_1,\beta_1,\gamma_2,\beta_2) =\nonumber\\ &g_1(\gamma_1,\beta_1+\beta_2)+g_1(\gamma_1,\beta_1)\cos^n\beta_2(e^{-i \gamma_2}-1).
\end{align}
To find the parameters that maximize overlap  $F_2(\gamma_1,\beta_1,\gamma_2,\beta_2)$, we set the gradients to zero and obtain a set of four equations. Even though in this case the variables do not separate, for $n\to\infty$ solutions behave as:
\begin{equation}
    n\beta_i\to \pi,  \gamma_i\to\pi.
    \label{guess}
\end{equation} 

\begin{figure}[!tbh]
 \begin{minipage}[b]{\linewidth}
   \centerline{\includegraphics[clip=true,width=3.2in]{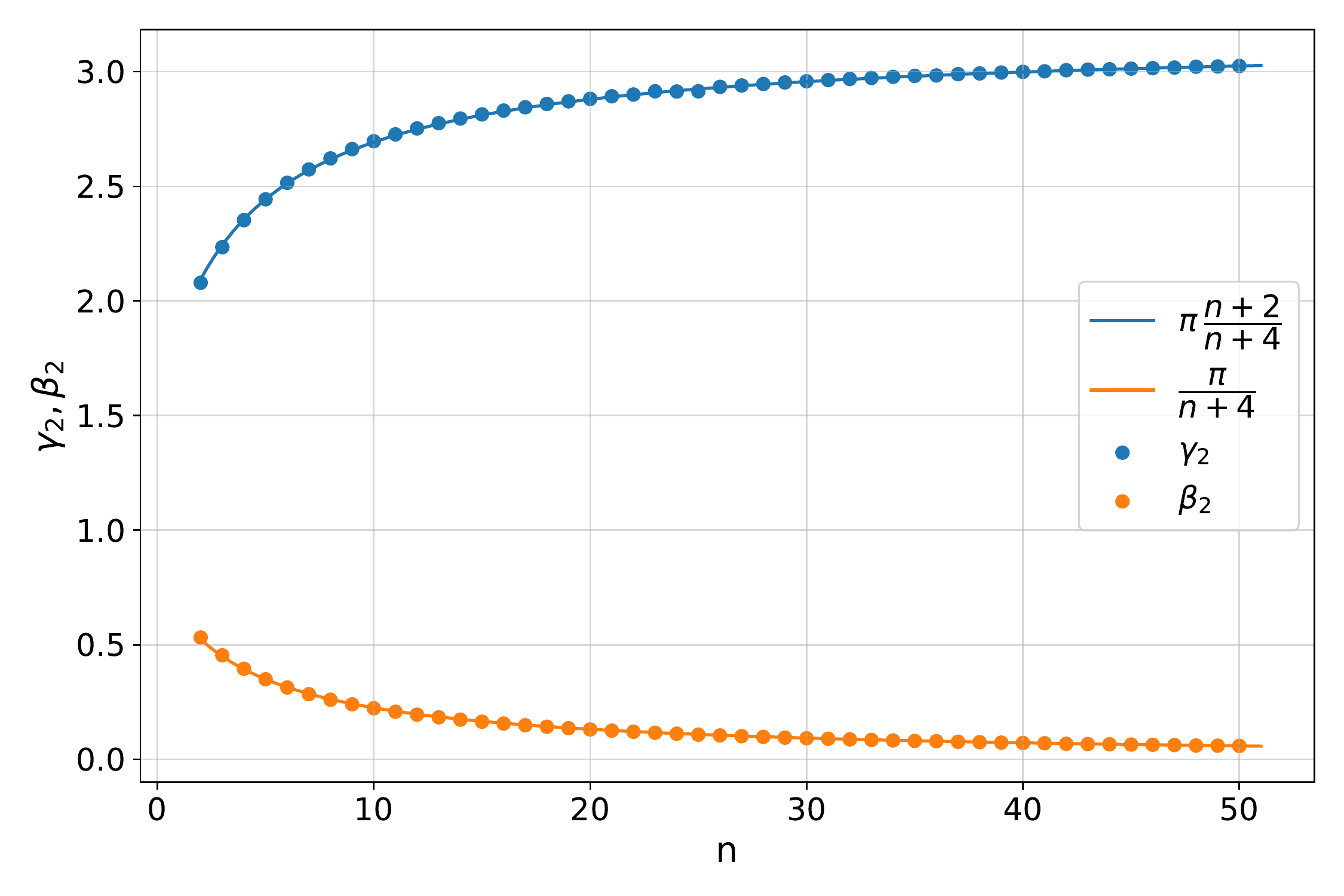}} (a)
   \end{minipage}
   \begin{minipage}[b]{\linewidth}
   \centerline{\includegraphics[clip=true,width=3.2in]{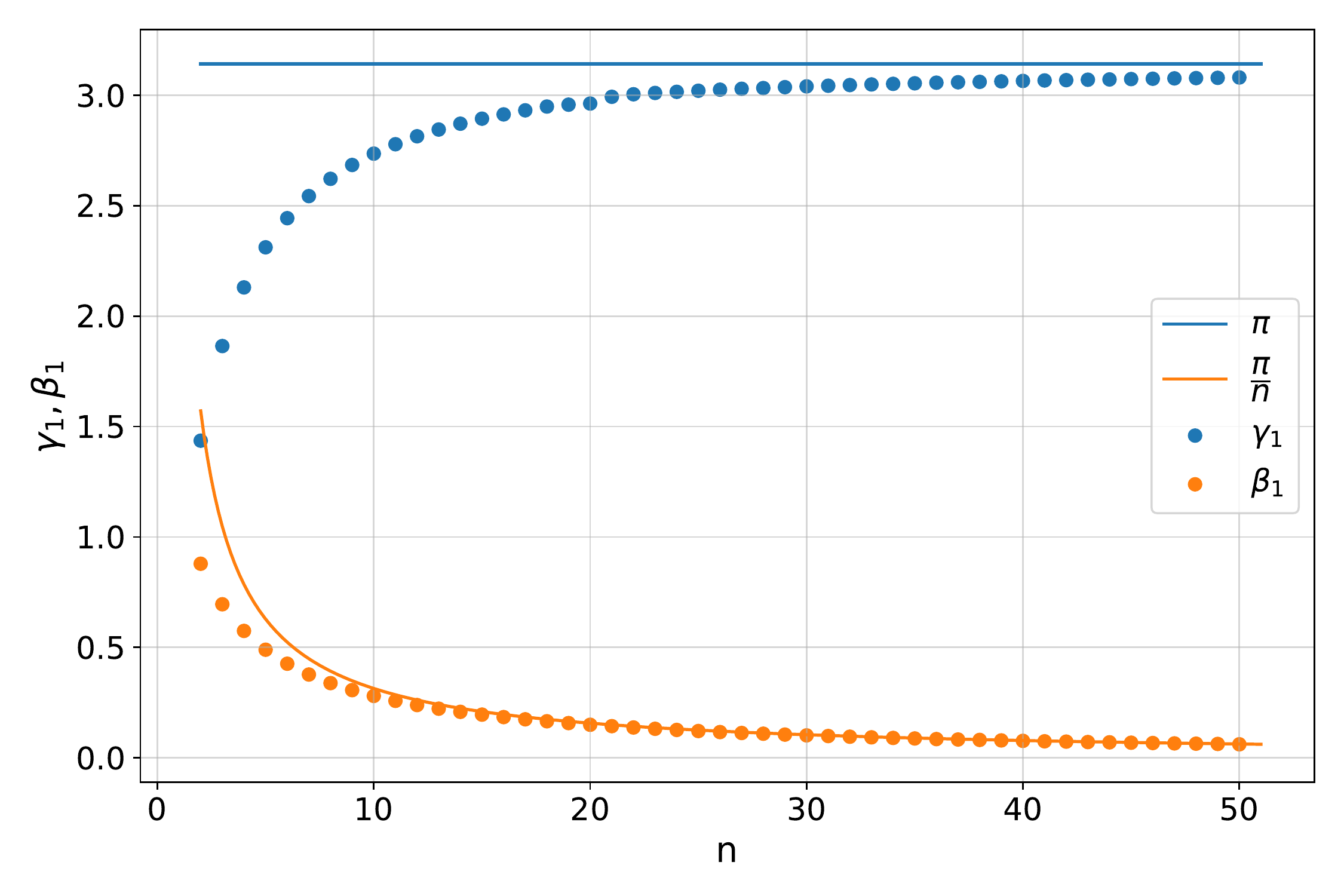}} (b)
   \end{minipage}
   \caption{Comparison of analytical and numerical results for optimal parameters (a) $\beta_2$ (orange), $\gamma_2$ (blue) and (b) $\beta_1$ (orange), $\gamma_1$ (blue), with respect to the number of qubits $n$. Dots represent numerical solutions to the maximization problem (\ref{objfun}). }
   \label{2_angles}
 \label{helix}
 \end{figure}

Assuming $n\gg1$ and corrections to be of the next order in $1/n$, we again search for parameters as $\beta_i=\dfrac{\pi}{n}+\delta\beta_i$ and $\gamma=\pi+\delta\gamma_i$ and obtain
\begin{align}
    \beta_2&=\dfrac{\pi}{n}-\dfrac{4\pi}{n^2}+\mathcal{O}(n^{-3})
    \label{2_beta_2},
    \\
    \gamma_2&=\pi-\dfrac{2\pi}{n}+\mathcal{O}(n^{-2})
    \label{2_gamma_2},
    \\
    \beta_1&=\dfrac{\pi}{n}+\mathcal{O}(n^{-3})
    \label{2_beta_1},
    \\
    \gamma_1&=\pi+\mathcal{O}(n^{-2}).
    \label{2_gamma_1}
\end{align}
It is seen that parameters $\beta_2$ and $\gamma_2$ behave in the same way as for the case $p=1$ in \eqref{beta} and \eqref{gamma}. Moreover, expressions $\beta_2=\dfrac{\pi}{n+4}$ and $\gamma_2 = \pi\dfrac{n+2}{n+4}$, which 
match our analytical solution for large $n$, also fall within optimal approximation for $n\sim1$ (see Figure ~\ref{2_angles}).

The corrections to the parameters $\beta_1$ and $\gamma_1$ turn out to be of the next order in $1/n$ and so we do not calculate them explicitly. In comparison to optimal parameters recovered numerically, the analytical predictions \eqref{2_beta_1} match optimal parameters beyond $n\sim 10$ (see Figure ~\ref{2_angles}). 

Note that for large $n$, corrections to parameters $\beta_1$ and $\gamma_1$ are of third and second order respectively. But for any finite region in $n$, parameters are well approximated by the functions 
\begin{equation}
    \beta = \dfrac{\pi}{a_1n+a_2},~\gamma = b_1\pi-b_2\beta,
    \label{fit_func}
\end{equation}
where the fitting constants $a_{1,2}$ and $b_{1,2}$ are region specific.

Based on our results, \eqref{2_beta_2}-\eqref{2_gamma_1}, we establish parameter concentration at $p=2$:
\begin{align}
    \abs{\bm\beta_{n+1}-\bm\beta_{n}}^2+\abs{\bm\gamma_{n+1}-\bm\gamma_{n}}^2=\mathcal{O}\left(\dfrac{1}{n^4}\right).
    \label{2_layer_concentration}
\end{align}

\subsection{Parameter concentration for $p\ge3$}
We have numerical evidence (up to $p=5$ and $17$ qubits) that for higher depths, optimal parameters also behave as \eqref{guess} for large $n$. Therefore, one might assume it to be a general feature. In order to calculate corrections for higher depth, the procedure described in the previous subsection can be used. However, a more straightforward approach is to Taylor-expand the overlap function around low-order solutions \eqref{guess} up to second order in $1/n$. This simplified expression is a quadratic form and thus can be maximized to obtain corrections. 

\begin{figure}[! tbh]
   \centerline{\includegraphics[clip=true,width=3.2in]{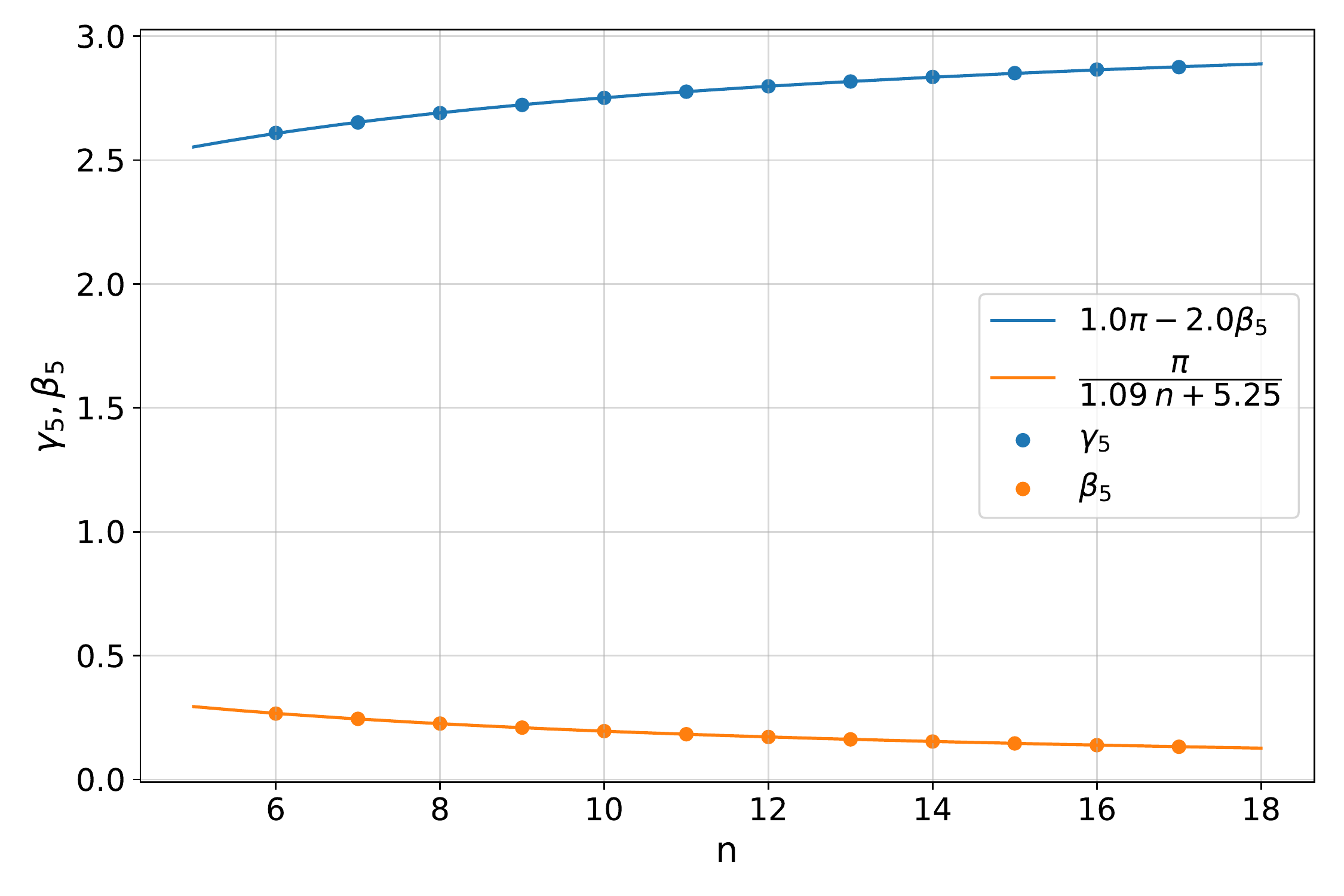}}
       \caption{Comparison of numerical results and our fitting for optimal parameters $\beta_5$ (orange), $\gamma_5$ (blue) with respect to the number of qubits $n$. Dots represent parameters obtained numerically via maximization of (\ref{overlap}). }
 \label{5_layer_angles}
 \end{figure}

Below we present our numerical results for $p=5$. Figure \ref{5_layer_angles} demonstrates numerically calculated optimal parameters for the last layer. Optimal parameters at depth $1-4$ are fit according to \eqref{fit_func} and the corresponding fit-constants appear in table I. 

\begin{figure}[!tbh]
    \centering
    \includegraphics[width= \textwidth]{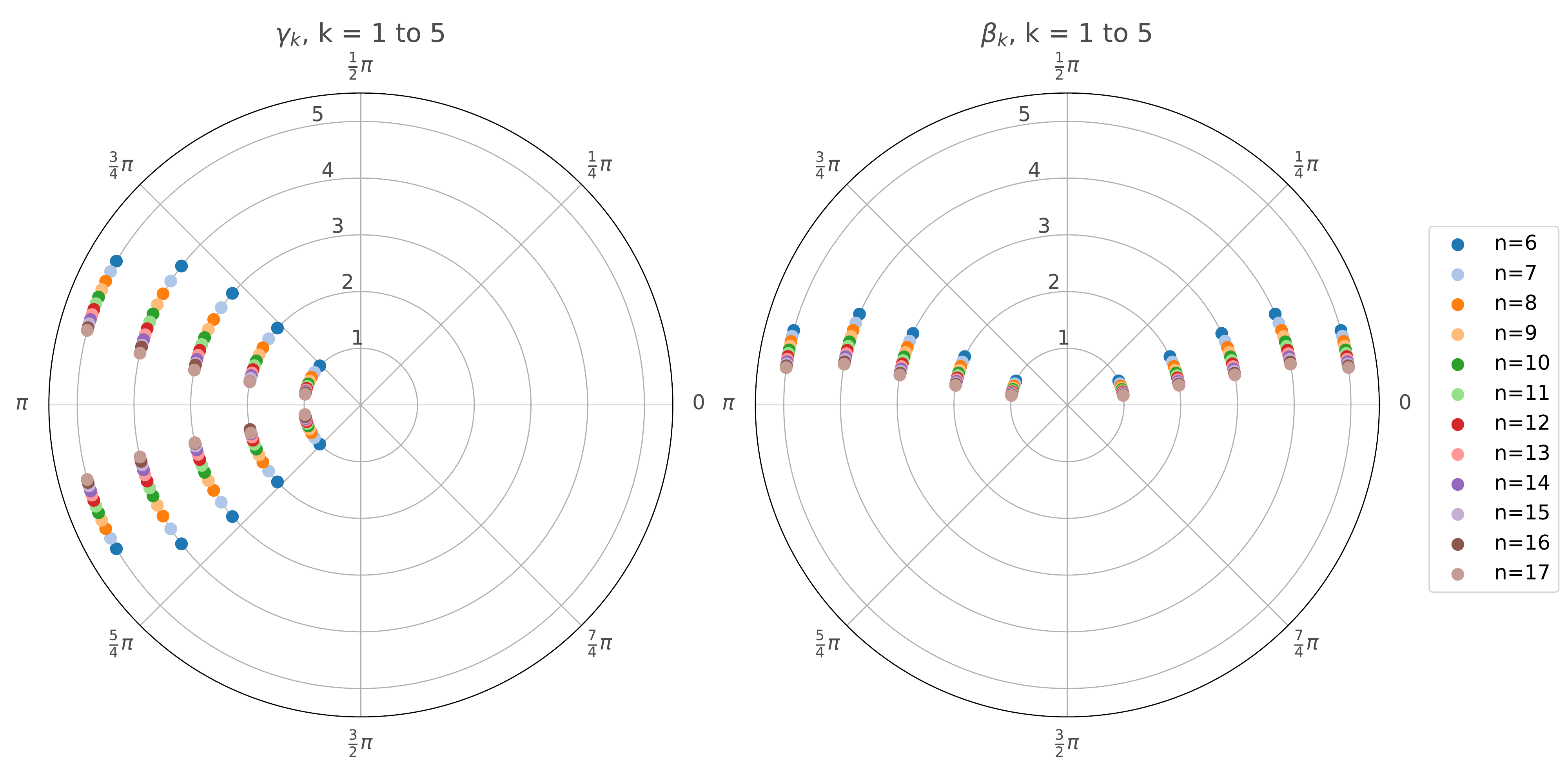}
    \caption{Parameter concentrations visualized for $p=5$ deep QAOA and varying number of qubits n. Two symmetric branches of optimal parameters are well distinguished. Parameter concentration is seen within each of the branches.}
    \label{circle}
\end{figure}

According to fitting curves which accurately describe the numerical data, parameter concentration is evident and is the same as in \eqref{2_layer_concentration}.
We also plot our numerical data in Figure \ref{circle} to visually illustrate the phenomena of parameter concentration.

\begin{table}[! tbh]
\centering
\begin{tabular}{|p{25pt}|p{25pt}|p{25pt}|p{25pt}|p{25pt}|}
\hline
&$a_1$&$a_2$&$b_1$&$b_2$\\
\hline
$\beta_1$,$\gamma_1$&$1.04$&$0.92$&$1.06$&$2.07$\\
\hline
$\beta_2$,$\gamma_2$&$0.98$&$1.23$&$1.05$&$2.04$\\
\hline
$\beta_3$,$\gamma_3$&$0.94$&$1.58$&$1.05$&$1.96$\\
\hline
$\beta_4$,$\gamma_4$&$0.88$&$2.32$&$1.03$&$1.83$\\
\hline
$\beta_5$,$\gamma_5$&$1.09$&$5.25$&$1.0$&$2.0$\\
 \hline
\end{tabular}
\label{fit_const}
\caption{Parameters $a_{1,2}$ and $b_{1,2}$ according to fitting functions (\ref{fit_func}).}
\end{table}

\section{Discussion}
\label{discussion}

In present work we provide a rigorous definition of parameter concentrations and demonstrate it for the variational state preparation. The definition is motivated by how this effect can be leveraged for efficient training. However, different approaches claiming concentrations appear in the QAOA literature, yet our  results have a clear distinction. Specifically \cite{Farhi2019a} analytically addresses what we call \textit{instance concentration} in the case of the Sherrington-Kirkpatrick model. Here, one finds that the variance in objective function value vanishes in the infinite size limit ($n \to \infty$), and therefore, QAOA  becomes instance independent. However, the result alone neither predict nor address the behavior of optimal parameters.

Numerically in \cite{Streif2020} it is seen that the optimal parameters at each depth $p$ distribute over a small range when considering randomly generated MAX-CUT instances on 3-regular graphs. Moreover, this distribution became narrower as the system size increased. The authors in \cite{Streif2020} explain their numerical observation via \textit{reverse causal cone} i.e.~the subgraphs effectively contributing to the objective function when taken over a particular edge. Since in the limit $n \to \infty$, the likely subgraphs are trees, optimal parameters for QAOA on such subgraphs become optimal for the entire graph. However, the scaling behavior (if any) for the optimal parameters towards the infinite $n$ limit still remain lacking. 

Although in our work only a unique set of optimal parameters is addressed at each $n$, their scaling behavior is fully understood due to our analytical result. Furthermore, we expect our results to be applicable in more general settings which would imply that distribution of optimal parameters with respect to instances are only slightly sensitive if $n$ is large. Such an implication can be leveraged to reduce the training  cost of finding optimal parameters. 

In particular, our observed concentrations scale as  $\mathcal{O}\left(n^{-4}\right)$, this implies optimal parameters also have a limit as $n\to \infty$. Therefore, one can train on a finite fraction $w\ll n$ qubits and perform a polynomially restricted training over optimal parameters at $w$ qubits to recover optimal parameters for $n$ qubits.
However, to fully exploit this approach further investigation is needed. 

Note that if parameters concentrate as $\mathcal{O}\left(n^{-l}\right)$ with $l\le2$, optimal parameters may not approach any limit. In this case partial training at $w$ qubits may not guarantee optimality for $n$ qubits. However the existence of such cases remains to be observed. 
\section*{Acknowledgements}
The authors acknowledge support from the research project, {\it Leading Research Center on Quantum Computing} (agreement No.~014/20).
\onecolumngrid

\bibliography{refs}
\bibliographystyle{unsrt}
\end{document}